# Low temperature tunneling current enhancement in silicide/Si Schottky contacts with nanoscale barrier width

Nicolas Reckinger,[1,a)] Xiaohui Tang (唐晓慧),[1] Emmanuel Dubois,[2] Guilhem Larrieu,[2,b)] Denis Flandre,[1] Jean-Pierre Raskin,[1] and Aryan Afzalian[1]

[1]*ICTEAM, Université catholique de Louvain, Place du Levant 3, 1348 Louvain-la-Neuve, Belgium*
[2]*IEMN UMR/CNRS 8520, Avenue Poincaré, BP 60069, 59652 Villeneuve d'Ascq Cedex, France*
[a)]Electronic mail : nicolas.reckinger@uclouvain.be.
[b)]Present address: LAAS CNRS, Université de Toulouse, 7, Avenue du Colonel Roche, 31077 Toulouse Cedex, France.

The low temperature electrical behavior of adjacent silicide/Si Schottky contacts with or without dopant segregation is investigated. The electrical characteristics are very well modeled by thermionic-field emission for non-segregated contacts separated by micrometer-sized gaps. Still, an excess of current occurs at low temperature for short contact separations or dopant-segregated contacts when the voltage applied to the device is sufficiently high. From two-dimensional self-consistent non-equilibrium Green's function simulations, the dependence of the Schottky barrier profile on the applied voltage, unaccounted for in usual thermionic-field emission models, is found to be the source of this deviation.

Metal oxide-semiconductor field-effect transistors (MOSFETs) with heavily doped source/drain (S/D) contacts face several technological difficulties when miniaturization below the 32 nm node is considered, e.g. for limiting short-channel effects arising from lateral diffusion of dopants in the channel during activation, and for reaching low sheet and contact resistances with ultra shallow junctions.[1] In this context, the replacement of conventional high doping S/D's by the Schottky barrier (SB) MOSFET architecture, where S/D contacts are metallic, was suggested.[2,3,4] Band-edge silicides, like rare-earth silicides[5,6] ($ErSi_{2-x}$, $YbSi_{2-x}$) for *n*-Si and PtSi[7,8] for *p*-Si, have been considered as potential candidates for SBMOSFET contacts since they achieve low SB heights (SBH) with electrons[9,10] ($\Phi_{Bn} \approx 0.3\,\text{eV}$) and holes[11] ($\Phi_{Bp} \approx 0.15\,\text{eV}$), respectively. However, such SBHs are still too high to compete with conventional MOSFETs in terms of on- and off-currents. As a solution, the dopant segregation (DS) concept, introduced by Thornton,[12] was recently revived by Kinoshita *et al.*.[13] DS consists in introducing a thin dopant layer at the silicide/Si interface to modulate the SBH. That layer promotes current injection by tunneling and results in a very low effective SBH ($\Phi_{Bn}^{\text{eff}}$).[14]

To investigate how DS affects the SBH, and more specifically parameters like the implantation dose, the implantation energy or the annealing temperature, accurate extraction of very low SBHs must be performed. Dubois and Larrieu developed an interesting solution to that issue. It consists of extracting the SBH from comparison of measured and modeled temperature-dependent current-voltage (*I-V*) characteristics of two Schottky contacts separated by a Si resistance. Current transport through the Schottky diodes was implemented according to the thermionic-field emission model of Crowell and Rideout (C&R).[15] The method was successfully applied to the precise SBH determination for both PtSi and $ErSi_{2-x}$.

In the present letter, we report strong deviations from the expected low temperature electrical behavior as predicted by C&R. We propose to elucidate the physical mechanism at the source of these deviations. The departure from the model could very probably impinge on a correct extraction of the effective SBH of segregated Schottky contacts.

The two-contact structure used for the electrical studies is pictured in Fig. 1, with $L_{Si}$ and $R_{Si}$ the length and the resistance of the Si gap, respectively. For $ErSi_{2-x}$, both segregated and non-segregated contacts to *n*-Si are investigated for long Si gaps ($L_{Si} \gg 1\,\mu\text{m}$). Both long and short ($L_{Si} \ll 1\,\mu\text{m}$) non-segregated PtSi/*p*-Si two-contact devices are also taken into consideration.

Since the deviations are not evidenced in the case efficient DS[16] (at least in the chosen measurement temperature range), we focus on purpose on $ErSi_{2-x}$/*n*-Si contacts implanted at low doses, exhibiting no or moderate SBH reduction. More interestingly, a dramatic disagreement appears at low temperature between the experimental data and C&R's model. This is illustrated in Fig. 2(a) displaying the experimental Arrhenius plots for three different kinds of $ErSi_{2-x}$/*n*-Si contact pairs: (i) non-segregated ($\Phi_{Bn} \approx 0.3\,\text{eV}$, blue star markers, sample 1), (ii) segregated but without effective SBH reduction ($\Phi_{Bn}^{\text{eff}} \approx 0.3\,\text{eV}$, green diamond markers, sample 2), and (iii) segregated with a moderate SBH reduction ($\Phi_{Bn}^{eff} \approx 0.2$ eV, brown circle markers, sample 3). The voltage is swept between 0.1 and 1 V with a step of 0.15 V. The corresponding modeled curves are also displayed in full lines of the same color. A typical Arrhenius plot features two distinct regions separated by a transition temperature: the ohmic region where $I$ is limited by $R_{Si}$ (positive slope) and the Schottky region where $I$ is limited by the Schottky contacts (negative slope). The temperature at which the transition occurs is indicative of the SBH and is used as criterion for data fit. C&R's model predicts an exponential drop of $I$ upon decreasing temperature in the Schottky regime, as seen for sample 1. We can see that the fit is excellent for sample 1 at all temperatures while it diverges from the experiment below the transition temperature (230 K) for sample 2 (same fit as sample 1 in the Schottky zone). The effect is even stronger for sample 3, below 150 K. The current do decrease below 150 K and the ohmic-to-Schottky transition occurs but the drop is far from exponential. In Fig. 2(b), the disparity is even better highlighted comparing the $I$-$V$ characteristics at 150 K of samples 1 and 2, respectively. Even though there is no apparent SBH reduction for sample 2, as testified by Fig. 2(a), the current injection is visibly somehow influenced by the dopant implantation. In addition, Figs. 3(a) and 3(b) illustrate that deviations can also be observed at low temperatures ($\leq$ 110 K) for non-segregated PtSi/$p$-Si devices with a short $L_{Si}$ (250 nm) as opposed to long $L_{Si}$ (25 µm).

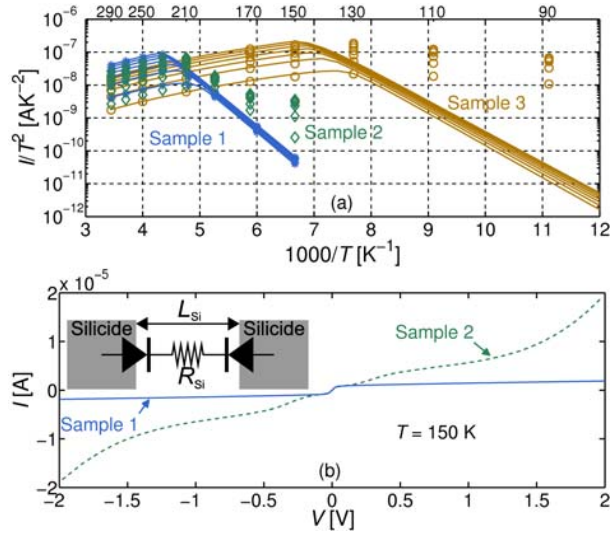

FIG. 1. (Color online) (a) Experimental (markers) and modeled (solid curves) Arrhenius plots for ErSi$_{2-x}$ two-contact structures: samples 1 (blue/stars), 2 (green/diamonds), and 3 (brown/circles). The fit in the Schottky region is the same for samples 1 and 2. (b) Corresponding measured $I$-$V$ characteristics at $T = 150$ K for samples 1 and 2.

In order to explain why $I$ in segregated or short devices increases with $V$ at low temperature, instead of saturating, we perform two-dimensional (2D) self-consistent non-equilibrium Green's function (NEGF) simulations. We use a coupled mode space approach assuming transport in the first subband only, an effective mass Hamiltonian, and ballistic transport approximation.[17] The SB is described as a contact potential.[18] For the DS case, the segregated region in Si is assumed uniformly doped with a concentration $N_{DS} = 10^{19}$ cm$^{-3}$ with a length $L_{DS} = 3$ nm.[19]

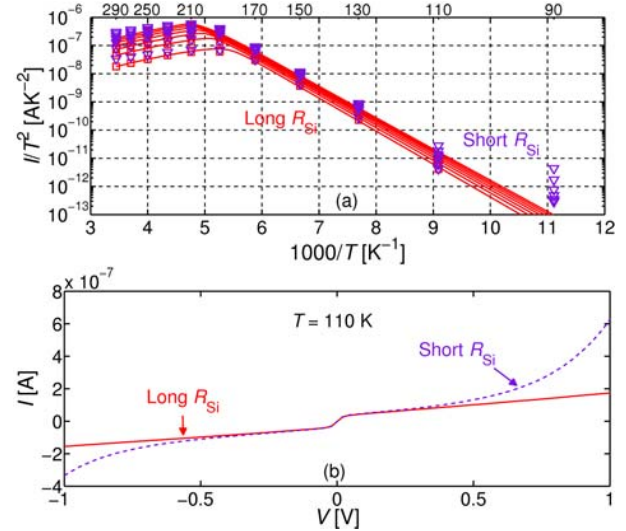

FIG. 3. (Color online) (a) Experimental (markers) and modeled (solid curves) Arrhenius plots for PtSi two-contact structures with long (red/squares) and short $R_{Si}$ (purple/triangles). (b) Corresponding measured $I$-$V$ characteristics at $T = 110$ K.

The current spectrum $J(E)$ in a Schottky contact is the result of a competition between occupation of electrons in the metal, mainly a Fermi-Dirac distribution $f_{FD}(E,T)$ (with $E$ the carrier energy above the Fermi level and $T$ the absolute temperature), and transmission probability of an electron through the barrier T($E$): $J(E) \approx f_{FD}(E,T) \times T(E)$. The first decreases exponentially above the metal Fermi level for increasing $E$ with a factor depending on the inverse of $T$: $f_{FD}(E,T) \propto \exp(-E/kT)$. On the other hand, T($E$) increases exponentially with a thinner energy-dependent barrier width $d(E)$ [T($E$) $\propto \exp(-d(E))$] and becomes equal to 1 for $d(E) = 0$ (i.e. above the SB). Moreover, it is weakly dependent on $T$. In turn, $d(E)$ decreases with $E$ at a rate that gets steeper with the donor doping $N_D$: $\partial d/\partial E \propto -N_D$. In consequence, $J(E)$ is lowered in energy with decreasing $T$, passing from thermionic emission over the SB at high $T$ to field emission (FE) through the SB at low $T$. In the temperature range considered here, transport essentially occurs via FE.

In Fig. 4, the energy band profile (conduction band minimum $E_C$) versus the transport direction $x$ of a short ($L_{Si}$ = 100 nm) segregated ErSi$_{2-x}$/$n$-Si device at T = 150 K is shown for increasing $V$. The excess of current at higher $V$ pictured in Fig. 2(b) can be linked to the much steeper band profile in the segregated region (inset 1 to Fig. 4) compared with the long non-segregated device, owing to a higher doping at the interface. In that case, the profile grows steeper with $V$, $d(E)$ decreases concomitantly, causing in turn an increase of T($E$) and $I$. This variation of the profile slope is not accounted for in C&R's model which supposes a fixed parabolic potential. As can be seen in inset 2 to Fig. 4, the corresponding $I$-$V$ curve renders very well, at least qualitatively, the behavior observed experimentally. The effect is the same for a long device [as featured in Fig. 2(a)] since it is only due to the interfacial energy band profile. In addition, as the non-saturation effect of $I$ with $V$ is correlated to the doping profile in the DS area, it could be used to electrically determine important parameters like $N_{DS}$ and $L_{DS}$, as a substitution or a complement to physical characterization methods.

mostly saturates with $V$ as expected from C&R's model.

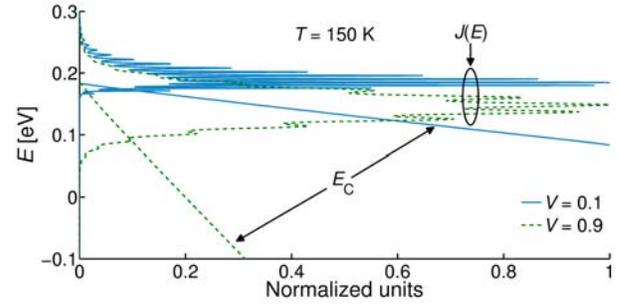

FIG. 4. (Color online) 2D self-consistent NEGF simulations of normalized $J(E)$ (rotated by 90°) versus $E$ and $E_c$ versus $x$/$L_{Si}$ for various $V$ at $T$ = 150 K for a short non-segregated ErSi$_{2-x}$/$n$-Si device. The current spectrum is normalized with respect to its maximum.

To conclude, we have reported low temperature disparities of the experimental $I$-$V$ characteristics of various two-contact Schottky structures compared with simulations relying on C&R's model. Based on 2D self-consistent NEGF simulations, it is highlighted that the dependence of the SB profile on $V$ results in an enhanced FE at low temperature. That profile modulation should be considered for a proper extraction of the SBH of dopant-segregated Schottky contacts and could prove useful to get a better insight into the dopant profile at the silicide/Si interface.

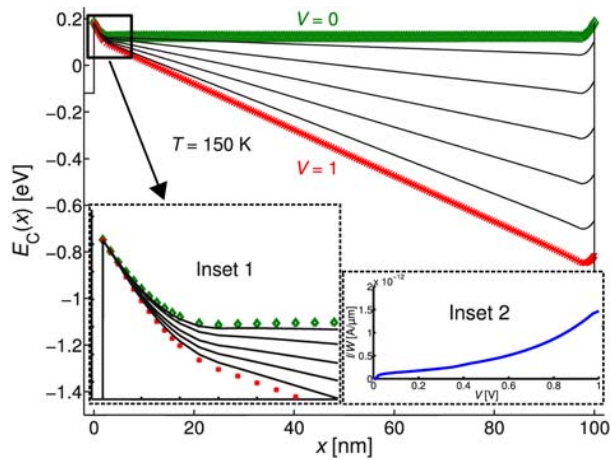

FIG. 3. (Color online) Energy band profile of a short segregated ErSi$_{2-x}$/$n$-Si device at $T$ = 150 K ($L_{Si}$ = 100 nm) given by 2D self-consistent NEGF simulations. Inset 1: zoom on the SB profile. Inset 2: corresponding $I$-$V$ characteristics.

In the case of a short non-segregated device, a similar non-saturation effect can also be observed at sufficiently low $T$, as shown in Fig. 3(b). The reason is that the rate of variation with $V$ of the band profile increases compared to a long device because the slope of the energy band versus $x$ is related to $V$/$L_{Si}$. In consequence, like in the segregated device, for a given $T$, a $V$ increment causes the band profile to grow more abruptly. Therefore the portion of $J(E)$ under the SB is significantly enhanced (see Fig. 5), which leads to an excess of current. In a long non-segregated device, however, the profile variation upon $V$ is negligible. The current spectrum is mostly determined by $T$, and $I$

This work was supported by the European Commission through the NANOSIL network of excellence (FP7-IST-NoE-216171) and by FRS-FNRS Belgium.